%% file: DPF2019_EmrahTiras.tex
\documentclass[11pt]{article}
\usepackage[margin=1in]{geometry}
\usepackage{graphicx}

\input econfmacros.tex 

\def\Title#1{\begin{center} {\Large {\bf #1} } \end{center}}
\def\Author#1{\begin{center} {\normalsize {\sc #1} } \end{center}}
\def\Institution#1{\begin{center} {\normalsize {\it #1} } \end{center}}
\def\Abstract#1{\noindent {\normalsize {\bf Abstract:} {\normalfont #1}}}
\def\Conference{\vspace{4mm}\begin{raggedright} {\normalsize {\it Talk presented at the 2019 Meeting of the Division of Particles and Fields of the American Physical Society (DPF2019), July 29--August 2, 2019, Northeastern University, Boston, C1907293.} } \end{raggedright}\vspace{4mm}}

\begin{document}

%
%

\Title{Detector R$\&$D for ANNIE and Future Neutrino Experiments}

\Author{Emrah Tiras}

\Institution{on behalf of ANNIE Collaboration\\Department of Physics and Astronomy \\ Iowa State University, Ames, IA, 50011, USA}

\Abstract{The Accelerator Neutrino Neutron Interaction Experiment (ANNIE) is designed to serve as a test bed for new detector technologies in future water and liquid scintillator based neutrino experiments. Located on the Booster Neutrino Beam at Fermilab, ANNIE will be the first gadolinium-loaded water Cherenkov detector on a neutrino beam and will provide high statistics measurements of neutron yields from neutrino interactions in water. It is also the first particle-physics application of the new photosensor technology: Large Area Picosecond Photodetectors (LAPPDs). With single photon time resolutions of roughly 50 psec and mm-level imaging capabilities, LAPPDs bring considerable new capabilities for neutrino reconstruction in Cherenkov and scintillator detectors. Leveraging this technology to make detailed neutrino measurements, ANNIE will serve as a first demonstration of their impact on physics. In addition to LAPPDs, the ANNIE R$\&$D program will likely explore other new technologies such as the addition of water-based liquid scintillator. The ANNIE Phase II detector is currently under construction and will start to take data in the summer of 2019. In this talk, I will present on the ANNIE detector R$\&$D program and its relevance to current and future neutrino experiments.}

\Conference{}

\section{Introduction}

The Accelerator Neutrino Neutron Interaction Experiment (ANNIE) is a water Cherenkov neutrino experiment located 100 meters downstream from the target of the Booster Neutrino Beam (BNB) at Fermilab. The experiment has three major detector components; a 26-ton of Gadolinium (Gd) loaded water Cherenkov detector, Muon Range Detector (MRD) and Front Muon Veto (FMV). During Phase I, ANNIE successfully built and operated a partially instrumented implementation of the full detector in order to measure cosmic and beam induced backgrounds. We found that beam induced neutron background is acceptable and it can be mitigated by 40-50 cm of water shielding around the fiducial volume \cite{Back:2017kfo}. A publication detailing this study is currently under collaboration review.

In the summer of 2019, the full implementation of the ANNIE physics phase (Phase II) detector was completed and is now under commissioning. Data taking is planned to begin with the turn on of the BNB, nominally at the end of October 2019. ANNIE physics phase aims to realize its main physics and R\&D measurements: (1) a unique hadron production measurement of neutrino-nucleus interactions, focusing specifically on neutron yield; and (2) an R\&D effort focused on using new photo-detector technology and key target material enhancements to enable advanced water-based neutrino detectors.

The primary physics goal of ANNIE is to study the multiplicity of final state neutrons from neutrino-nucleus interactions and charged current quasi-elastic (CCQE) cross section measurements in water. ANNIE provides a unique opportunity to study this physics in a controlled beam experiment in an energy range of 700 MeV, which is relevant to both atmospheric neutrinos and long baseline experiments. ANNIE's neutron physics program is highly complementary to similar measurements of proton multiplicity in the BNB using liquid argon time projection chambers (LAr-TPCs).

The primary technological goals are to demonstrate new detection technologies such as Large Area Picosecond Photodetectors (LAPPDs) \cite{Adams:2016tfm, TimingLAPPD} and neutron tagging in Gd-loaded water. ANNIE is the first application of LAPPDs, and we have been building and developing characterization tests and new reconstruction techniques over the last 2-3 years. Also, ANNIE has developed a low-cost water purification/circulation system for smaller scale Gd-water deployments. UC-Davis has developed a low-cost sulfate-loaded resin by using commercially available materials and this resin removes all the nitrates and free radicals from the water, and leaves only Gd-sulfate in the solution.  

\section{Physics Phase Detector Overview}

\begin{figure}[htb]
\centering
\includegraphics[scale=0.4]{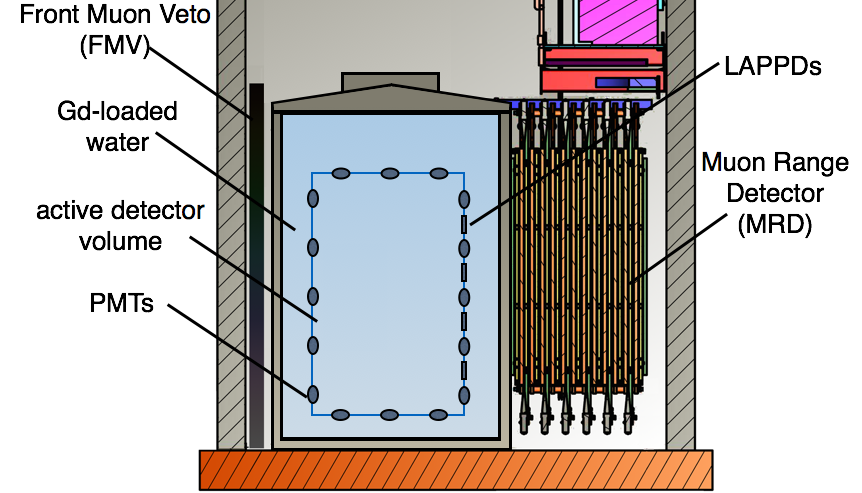}
\caption{A concept drawing of the ANNIE physics phase detector.}
\label{fig:detector}
\end{figure}

A concept drawing of the ANNIE physics phase (Phase II) detector is shown in Fig \ref{fig:detector}. The main volume consists of an upright cylindrical tank (10 ft diameter x 13 ft tall), filled with 26 tons of Gadolinium- loaded water. The Gd enhances the neutron-capture cross section of the target and produces a detectable characteristic (8 MeV) gamma ray signal within a much shorter time frame (O(10)$\mu$s)  than that of hydrogen (O(100)$\mu$s). Neutrons that thermalize in the target will be detectable from the optical light produced by these gammas, collected by a mixture of 11-in, 10-in, and 8-in conventional photomultiplier tubes (PMTs) surrounding the active volume. Five Large Area Picosecond Photodetectors (LAPPDs) deployed on the downstream face of the tank will be used for detailed reconstruction of the Cherenkov light. Rock muons entering from outside of the detector will be tagged using muon paddles in the FMV. The MRD, recommissioned from the SciBooNE experiment, is used to reconstruct the direction and range-out energy of muons produced by charged current interactions inside the tank. 

\vspace{1 cm}

\begin{figure}[h]
\centering
\includegraphics[scale=0.45]{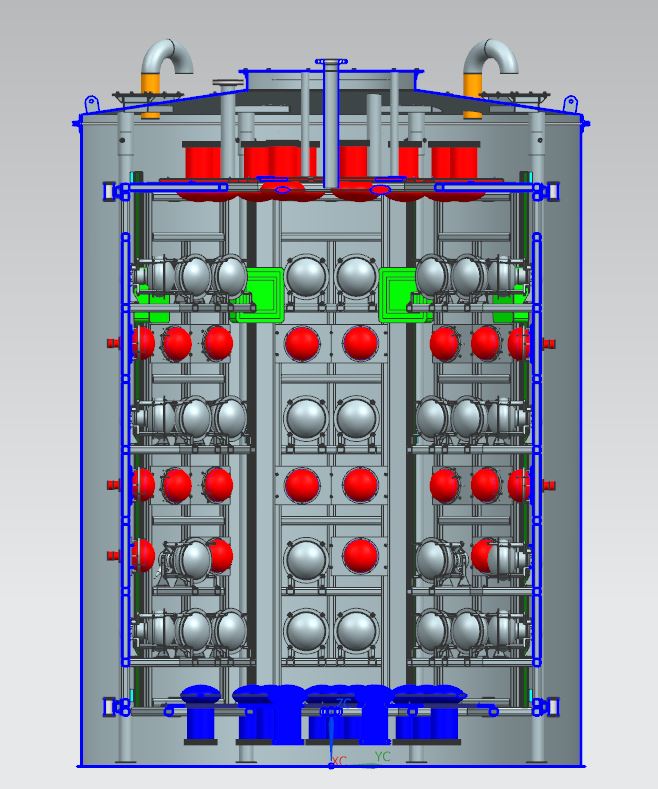}
\includegraphics[scale=0.222]{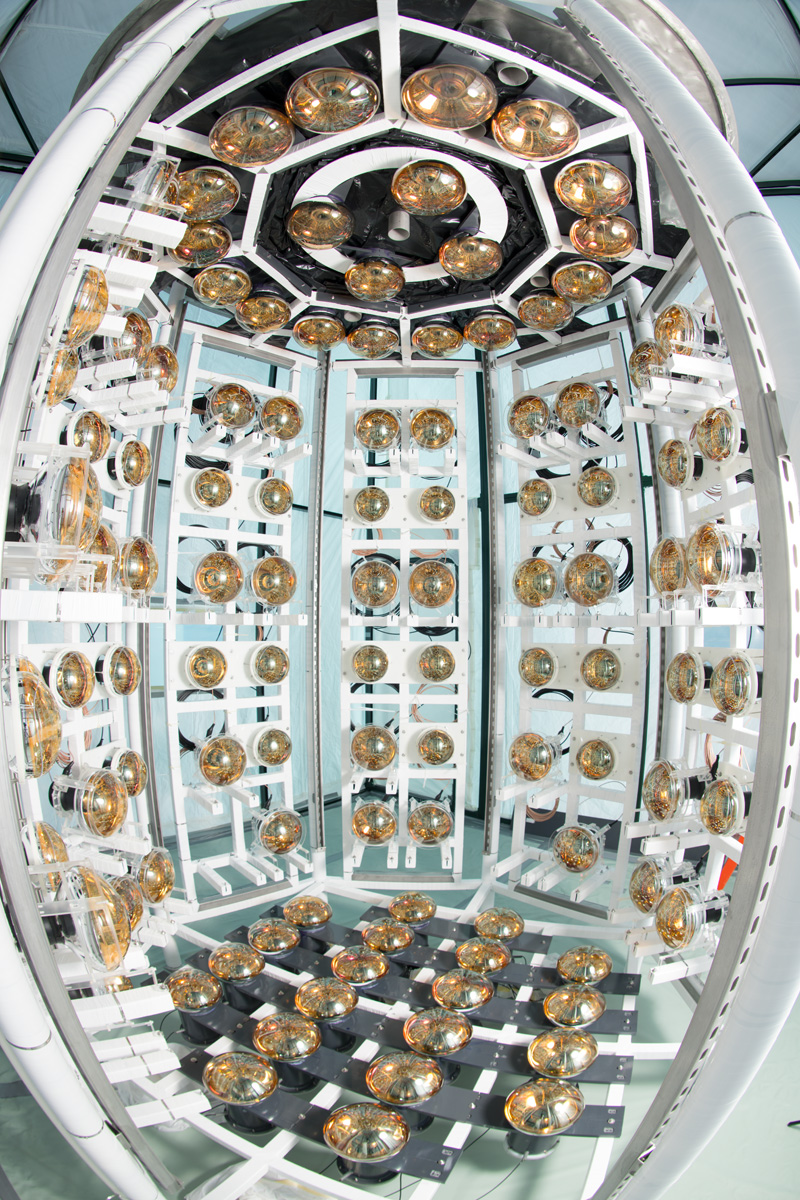}
\caption{The 3D design (left) and a picture of the inner structure (right) of the Phase II detector. }
\label{fig:innerstructure}
\end{figure}

The PMTs and LAPPDs are supported by a 304-stainless steel, self-supporting, octagonal inner structure hanging from the lid of the tank. The lid and inner structure can be lowered into and removed from the tank. Since the inner structure is free standing, PMTs can be mounted onto the lattice in the open, removing difficulties involved in working in confined spaces. The Phase I inner structure was a simpler version to hold sixty 8-inch PMTs at the bottom only and the Phase II has a higher density lattice of steel extrusions on all sides in order to accommodate the higher number and isotropic distribution of Phase II PMTs and LAPPDs. The PMTs are attached to removable panels on the sides. This inner structure, pictured in Fig. \ref{fig:innerstructure} (right), is designed to hold as many as the 96 side-mounted PMTs, 20 upward facing PMTs, and 20 downward facing PMTs. This provided enough space to accommodate 132 conventional PMTs in ANNIE Phase II. Eight “mail-slots” on the lid of the tank, combined with tracks on the columns of the inner structure will enable in-situ deployment of LAPPDs in the target volume, even after water fill.

Several steps were taken to minimize the risk of rust forming on the steel. The Phase II inner structure was rebuilt, based on the original Phase I design, in order to mitigate risks associating with minor rust spots. Among those steps, the steel pipes and extrusions are to be pickle-passivated and inspected before construction. Also, sample pieces from each type of pipes and extrusions were tested in Gd-loaded water at UC-Davis. Care was taken during the welding process to preserve the integrity of the stainless and electro-polishing was performed on the welds and scratches on the inner structure afterwards. Then, the whole inner structure was wrapped in 2 inch Teflon tape to further protect the steel.  

Figure \ref{fig:innerstructure} (left) shows the 3D design of the inner structure with PMTs in different colors, which represents different types of PMTs. Twenty 10-inch Hamamatsu PMTs (Blue in Fig. \ref{fig:innerstructure} - left) from LUX experiment were mounted at the bottom of the inner structure. The sides were populated with a mixture of the  forty-eight 10-inch WATCHBOY \cite{Dazeley:2015uyd} and four 10-inch high-quantum-efficiency (HQE) WATCHMAN PMTs \cite{Askins:2015bmb} (both are Gray in Fig. \ref{fig:innerstructure} - left)  and the newly purchased forty 8-inch HQE Hamamatsu PMTs (Red in Fig. \ref{fig:innerstructure} - left). The top of the tank was built for holding twenty 11-inch HQE ETEL PMTs (also Red in Fig. \ref{fig:innerstructure} - left). With all the PMTs, we have about 20\% photocoverage in the tank.

Two PMT characterization facilities were built for the ANNIE experiment; at UC-Davis and Fermilab. We measured the gain, after-pulsing and cool-down time. Their gains are in between $10^6$-$10^7$ at optimum operating voltages and they are linear as a function of the applied voltage values. For most of the PMTs, there is a 10-15\% after-pulsing rate, which is acceptable. Almost all the PMTs cool down in about 30-40 minutes after they are left in a darkbox with no external light source. 

\begin{figure}[h]
\centering
\includegraphics[scale=0.33]{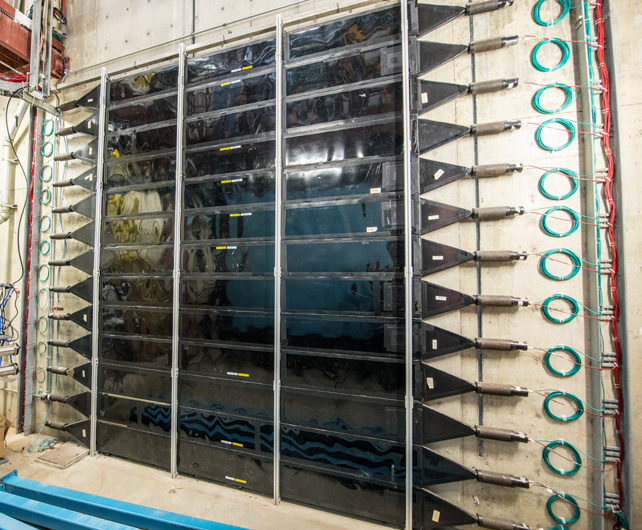}
\includegraphics[scale=0.34]{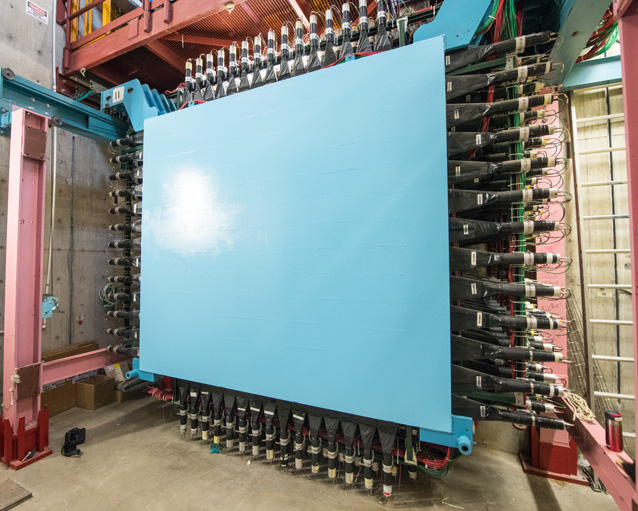}
\caption{A view of the Front Muon Veto (FMV) (left) and Muon Range Detector (MRD) (right).}
\label{fig:mrd}
\end{figure}

Figure \ref{fig:mrd} (left) shows the FMV, which consists of two overlapping layers of muon paddles, in total of 26 scintillating paddles instrumented with 2" PMTs, inherited from the CDF experiment. Figure \ref{fig:mrd} (right) shows the MRD which is an iron and plastic scintillator sandwich calorimeter, designed to range muons up to about 1 GeV and provide directional information and energy of the stopped muons. It was inherited from the SciBooNE experiment and left roughly 2/3 intact. The rest, 1/3 of the MRD has been reconstructed with refurbished scintillator paddles from the KTeV experiment.

The MRD has 11 alternating vertical (5) and horizontal (6) layers with 5 cm iron absorbers in between. The horizontal layers consist of 2 sets of 13 horizontal paddles in 6 layers, and a total of 156 channels. For the vertical layers, the first layer has 2 columns of 15 paddles, the second layer has 2 columns of 17 paddles,the third vertical layer has 2 columns of 13 paddles and the other two layers consist of 2 columns of 15 paddles. All together there are 306 scintillator paddles on the MRD, which are all read out with 2 inch PMTs. 64 paddles for the first two vertical layers refurbished from the KTeV experiment, and the rest of the 306 paddles were inherited from the SciBooNE experiment. Those 64 paddles from the KTeV experiment plus 8 paddles from the SciBooNE experiment (same with the rest of the MRD paddles) have been characterized. The efficiency of those scintillator paddles have been measured by cosmic muons as 3-fold coincidences as a function of hit positions at various voltage values, 1500-2200 V. We found that most of them have efficiency in between 80-90\%. 

%


\section{Large Area Picosecond Photo-Detectors (LAPPDs)}

Starting in December 2019, ANNIE will become the first application of LAPPDs in Gd-loaded water in a neutrino beam. LAPPDs are 20 cm x 20 cm flat panel, microchannel plate (MCP) based photodetectors with roughly 50 picosecond single photoelectron (sPE) time resolutions. They are capable of milimeter-scale spatial resolution, less than 1 cm across a strip, and less than 5 mm along a strip depending on the readout system. They have borosilicate glass window and multi-alkali ($K_2NaSb$) photocathode material. Each LAPPD has 28 strips and read out from both ends of the strips by two 30 channel ACDC cards attached to an analog pickup board on the back. Their gain characteristics are exceeding $10^7$, well comparable with the conventional PMTs. 

The time resolution of LAPPDs is a very powerful capability for photodetectors, for neutrino and in general particle physics experiments since they enable: 

\begin{itemize}
\item improved vertex reconstruction, 
\item improved ability to reconstruct overlapping events and tracks, 
\item improved ability to resolve the structure of EM showers, 
\item stronger handles for background rejection and energy reconstruction in higher energy beams.
\end{itemize}

LAPPDs are imaging detectors, able to resolve the positions of photon hits, rather than measuring total charge on a single pixel. This specification of LAPPDs allows the identification of multiple individual photons. 

Due to their fast timing and powerful imaging capabilities, they can be used for many applications in particle physics, nuclear physics, X-ray science, and medical imaging. In a recent study E. Angelico et.al. \cite{Angelico:2019gyi} showed that timing can be used to select different energy fluxes from a wide-band of neutrino beam. Lower energy neutrinos come from slower pions and they arrive at the detector later. Their suggestion is to increase the frequency of Fermilab radio-frequency (RF) by 10 times and rebunch the beam to make this effect exploitable. This will require precision time-stamping O(100) ps.  

They also have strong potential to separate between Cherenkov and Scintillation time, based on the arrival time of the light \cite{Aberle:2013jba}. T. Kaptanoglu et.al. in 2018 \cite{Kaptanoglu:2018sus} showed that chromatic separation by using dichroic filters makes the seperation of Cherenkov and scintillation light possible. LAPPDs can also be tested in plenoptic imaging applications, i.e. an application of J. Dalmasson et.al.'s \cite{Dalmasson:2017pow} study to focus different fluxes of scintillation light as a function of wavelength by recording intensity, color and directional information of the light. LAPPDs can be used in detector where multi-bounce optics are used to collect/track more Cherenkov photons for a through-going charged particle track. E. Oberla and H.J. Frisch tested a small prototype optical-TPC (OTPC) in 2015 at Fermilab Test Beam Facility and used microchannel plate PMTs (MCP-PMTs) with mirrors on the detector wall opposing the MCP-PMTs. They collected more Cherenkov photons by using this multi-bounce optics technique and measured the delay of the Cherenkov photons as 770 ps compared to the directional Cherenkov light \cite{Oberla:2015oha}. ANNIE will be able to demonstrate many of these techniques and new capabilities. The detector nominally has 5 LAPPDs but is built so that more can be added as the become available. Our most recent Monte Carlo results show an improvement from 40 cm to 20 cm (at 68\% CL) on position resolution with the addition of 5 LAPPDs to the conventional PMTs. 

\begin{figure}[h]
\centering
\includegraphics[scale=0.05]{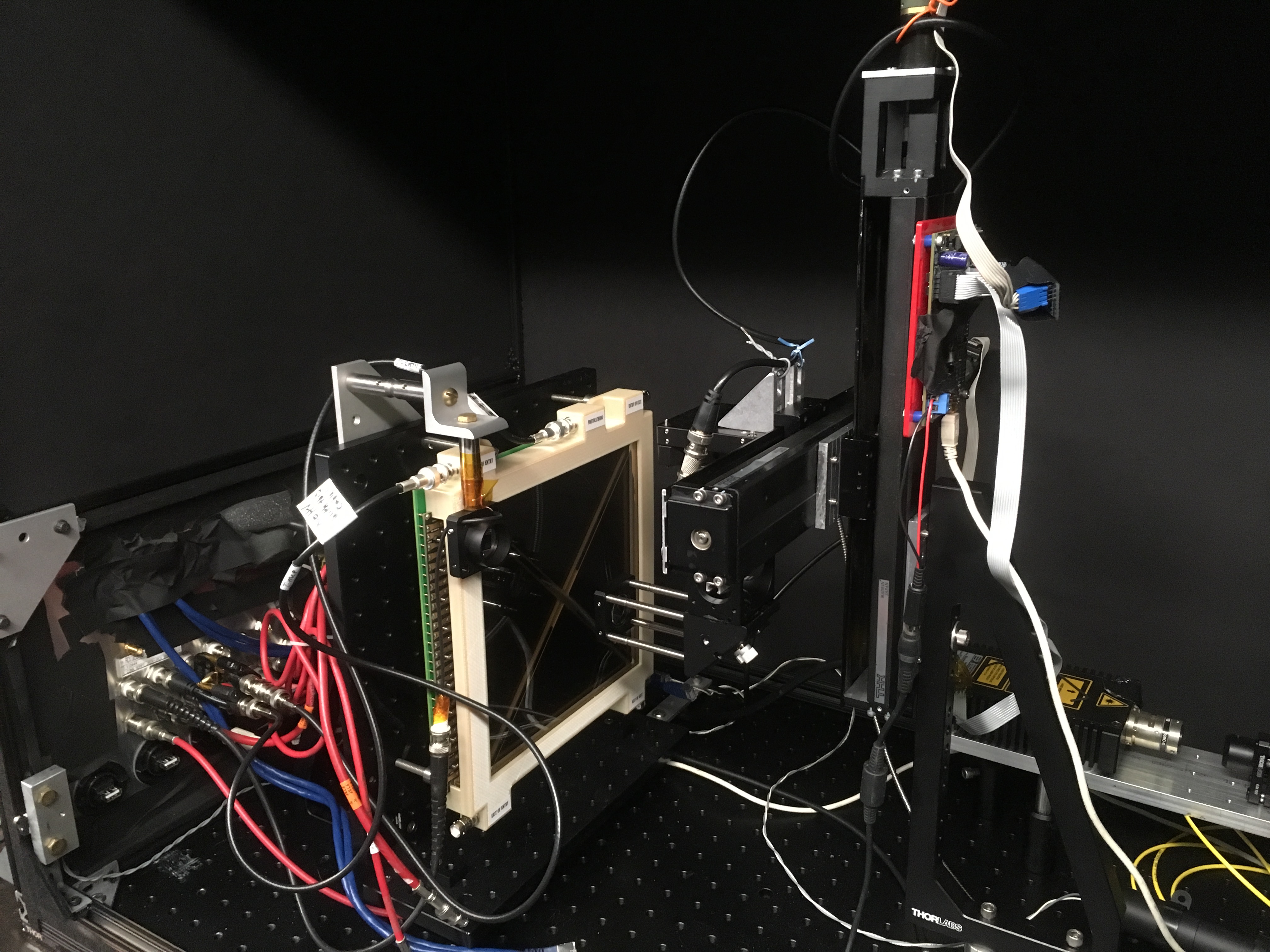}
\includegraphics[scale=0.66]{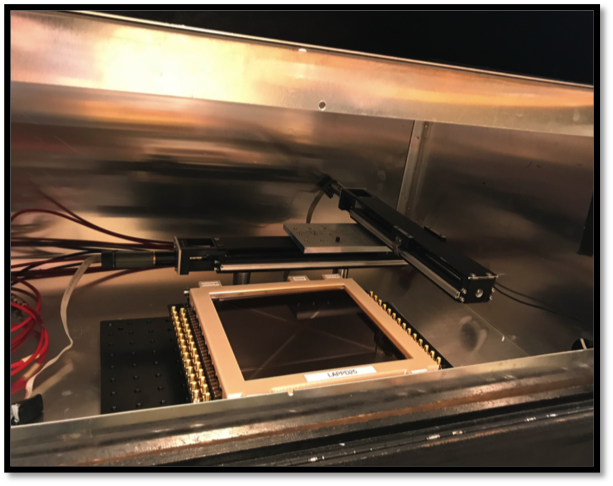}
\caption{Characterization laser test setup of LAPPDs at Fermilab (left) and at ISU (right)}
\label{fig:lappd}
\end{figure}

The ANNIE collaboration built two LAPPD characterization facilities, based on prior work by ANNIE collaborators \cite{Adams:2016tfm, Adams:2013nva, Minot:2018zmv}. The first one is at Iowa State University (ISU), Fig. \ref{fig:lappd}-right and the second one is at Fermilab, Fig. \ref{fig:lappd}-left.  

The characterization setup at Fermilab is centered in a front-access dark-box with dimensions of 36 in. (W) by 24 in. (D) by 32 in. (H), inside of which there are a mount for the LAPPD under test, a 50-ps-pulsed 405-nm diode laser, a 420-nm LED, and a 2D motion stage for delivering the light from these sources to any place on the detector. Light from both sources is delivered by means of optical fibers to beam-forming optics on the motion stages. Mounted on the backside of the LAPPD under test, there are two waveform sampling readout boards with a total of 60 channels at 10 Gsample/sec, based on the PSEC4 architecture developed at the University of Chicago. This dark box is instrumented with two pico-amperemeters, a relay switch matrix, two low-voltage power supplies, and four high-voltage power supplies, all of which are under computer control for fully automated test sequences. The motion stage and the PSEC4 electronics are also controlled by the computer. The software for integrating all instrumentation and running test sequences is open-source custom code \cite{Bernhard}.

Quantum-efficiency maps are obtained by first measuring the light reaching the detector with a NIST-traceable photodiode positioned on the perimeter of the LAPPD, and calibrating a reference photodiode in the optics head on the motion stages against it. Then, it takes dark-current measurements, and then scanning the optics head in, typically, a raster of 81 by 81 points (2.5 mm steps) over the photocathode while measuring photocurrent from it, as well as the current from the freshly calibrated reference photodiode.

\begin{figure}[h]
\centering
\includegraphics[scale=0.55]{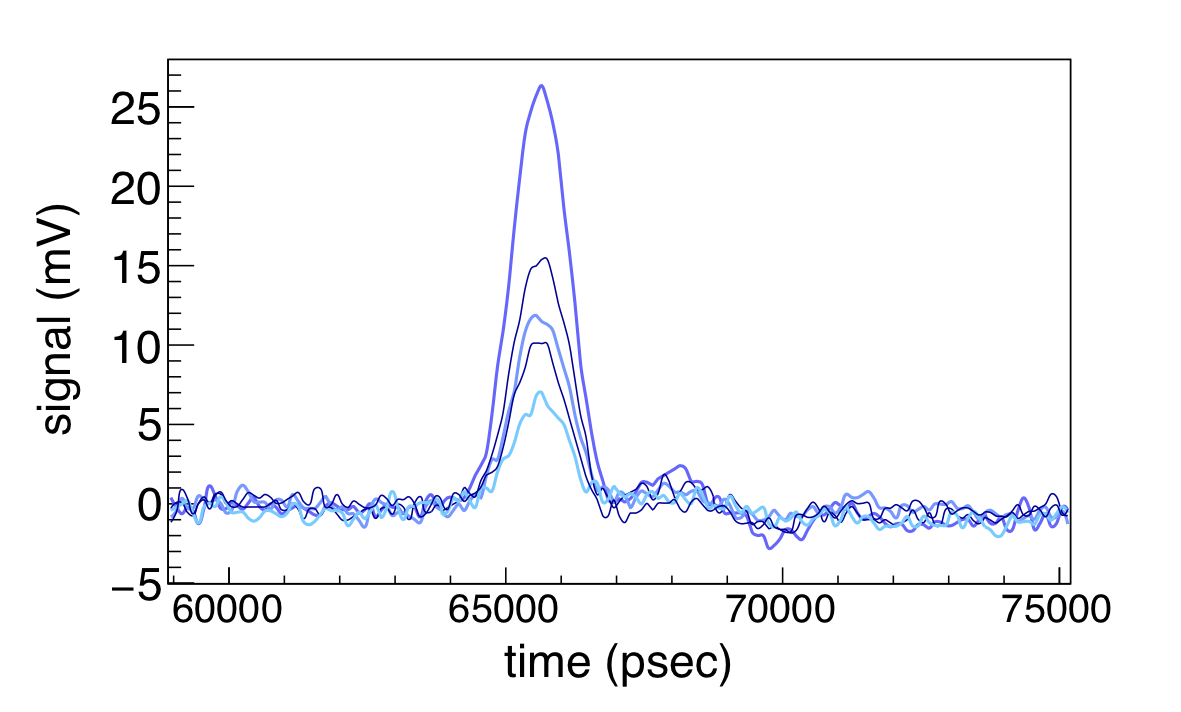}
\caption{Some typical sPE pulses from one of the LAPPDs.}
\label{fig:lappd_SEpulse}
\end{figure}

\begin{figure}[h]
\centering
\includegraphics[scale=0.5]{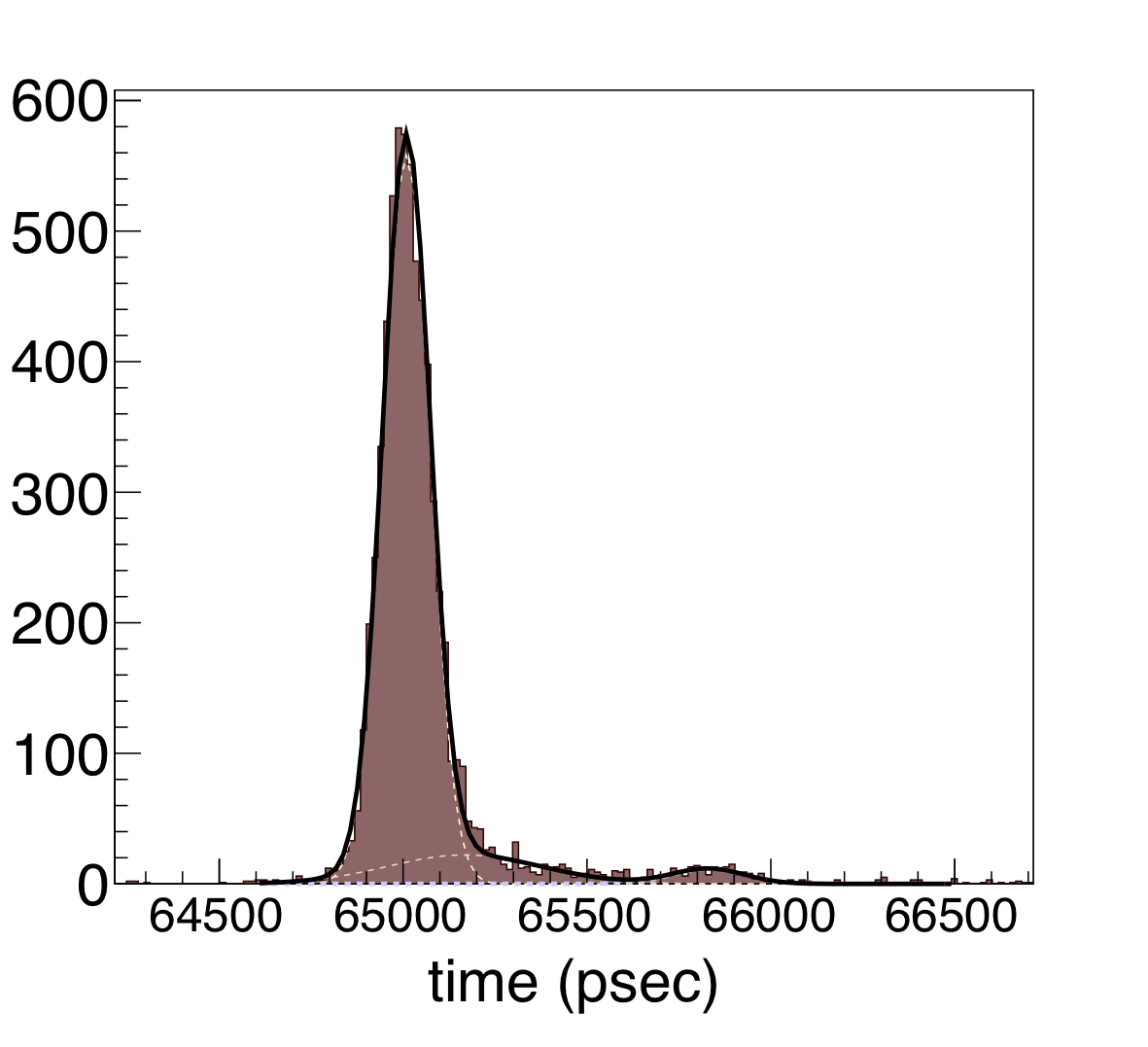}
\caption{Transit time spread of the same LAPPD, the distribution is fit with three Gausians.}
\label{fig:lappd_tts} 
\end{figure}

Simlarly, the sPE response of the LAPPD to laser pulses is tested by scanning the location with the motion stages, as well as the high voltages applied to the MCPs. Gain and timing calibration tests are still underway with rigorous analysis and simulation program. Figure \ref{fig:lappd_SEpulse} shows some typical sPE pulses from one of the LAPPDs (LAPPD \#25), the rise time of the pulses was measured as 850 picoseconds and full-width half maximum (FWHM) as 1.1 nanoseconds roughly. Figure \ref{fig:lappd_tts} shows the transit time spread of the same LAPPD, the time resolution is measured less than 60 picoseconds with less than 4\% of after-pulsing.    

In this paper we have reported on some sampling characterization results of one LAPPD only, but a very detailed characterization results from the five ANNIE LAPPDs will be published as a separate article soon. 

\section{Conclusion}

The Accelerator Neutrino Neutron Interaction Experiment (ANNIE) at Fermilab is on track to measure the neutron abundance and CC cross section measurements from neutrino interactions in a water Cherenkov detector. Measurements of final-state neutron multiplicity will improve our understanding of the complex, many-body dynamics of neutrino-nucleus interactions. This will provide a strong experimental handle for the systematic uncertainties of the neutrino energy reconstruction. To achieve this goal, ANNIE will deploy Large Area Picosecond Photodetectors (LAPPDs), which will enable significant improvement for vertex and track reconstruction and thus improve the energy resolution and background rejection capabilities. We have five LAPPDs ready on hand, and characterization tests are underway at Fermilab. They will be deployed in-situ by the end of the year and ANNIE will first demonstrate this novel technology in a neutrino beam. The flexibility and portability of ANNIE's detector gives a chance to expand its program easily as a testbed for new technologies such as Water-based Liquid Scintillator (WbLS). In addition, ANNIE’s low-cost water purification/circulation system can be used by similar scale deployments of Gd-water in future experiments.

\section*{Acknowledgements}

This work could not have been accomplished without personnel, facilities, and resources support of Fermi National Accelerator Laboratory, operated by Fermi Research Alliance, LLC under Contract No. DE-AC02-07CH11359 with the United States Department of Energy. The activities of the ANNIE experiment are supported by the U.S. Department of Energy, Office of Science, Office of High Energy Physics under contracts DE-SC0016326 and DE-SC0019214, together with Science and Technology Facilities Council and Scottish Universities Physics Alliance (United Kingdom). We also would like to thank to Evan Angelico from the University of Chicago and Dr. Bernhard Adams from Incom, Inc. for their tremendous help and contributions with the LAPPD testing setup at Fermilab.

\end{document}

%% file: econfmacros.tex
\def\beq{\begin{equation}}
\def\eeq#1{\label{#1}\end{equation}}
\def\eeqn{\end{equation}}

\def\beqa{\begin{eqnarray}}
\def\eeqa#1{\label{#1}\end{eqnarray}}
\def\eeqan{\end{eqnarray}}

\let\bar=\overbar

\def\Dslash{\not{\hbox{\kern-4pt $D$}}}
\def\dslash{\not{\hbox{\kern-2pt $\del$}}}

\def\msb{{\bar{\ssstyle M \kern -1pt S}}}

